\documentclass[reprint,aps,prb]{revtex4-2}

\usepackage{graphicx}  
\usepackage{subfigure}
\usepackage{multirow}
\usepackage{comment}
\usepackage{fancyhdr}
\usepackage{longtable}
\usepackage{parskip}
\usepackage[T1]{fontenc}
\usepackage{dcolumn}   
\usepackage{bm}        
\usepackage{amsfonts}  
\usepackage{amsmath}   
\usepackage{amssymb}   
\usepackage{siunitx}   
\usepackage{soul}
\usepackage{cancel}
\usepackage{hyperref}

\DeclareRobustCommand{\rchi}{{\mathpalette\irchi\relax}}
\newcommand{\irchi}[2]{\raisebox{\depth}{$#1\chi$}} 

\begin{document}

\title{Nonlinear magnetotransport in MoTe${}_2$}

\author{A.C.~Marx}
\email{a.c.marx.goncalves@rug.nl}
\author{H.~Jafari}
\author{E.K. Tekelenburg}
\author{M.A. Loi}
\author{J.~S{\l}awi\'nska}
\author{M.H.D.~Guimar\~aes}
\email{m.h.guimaraes@rug.nl}

\affiliation {Zernike Institute for Advanced Materials, University of Groningen, NL-9747AG Groningen, The Netherlands}

\begin{abstract}  

The shape of the Fermi surface influences many physical phenomena in materials and a growing interest in how the spin-dependent properties are related to the fermiology of crystals has surged. 
Recently, a novel current-dependent nonlinear magnetoresistance effect, known as bilinear magnetoelectric resistance (BMR), has been shown to be not only sensitive to the spin-texture in spin-polarized non-magnetic materials, but also dependent on the convexity of the Fermi surface in topological semimetals. 
In this paper, we show that the temperature dependence of the BMR signal strongly depends on the crystal axis of the semimetallic MoTe${}_2$. 
For the a-axis, the amplitude of the signal remains fairly constant, while for the b-axis it reverses sign at about \SI{100}{\kelvin}. We calculate the BMR efficiencies at \SI{10}{\kelvin} to be \(\rchi_{J}^{A}=\) \SI{173(3)}{\nano\meter\squared\per\tesla\per\ampere} and \(\rchi_{J}^{B}=\) \SI{-364(13)}{\nano\meter\squared\per\tesla\per\ampere} for the a- and b-axis, respectively, and we find that they are comparable to the efficiencies measured for WTe${}_2$. We use density functional theory calculations to compute the Fermi surfaces of both phases at different energy levels and we observe a change in convexity of the outer-most electron pocket as a function of the Fermi energy. Our results suggest that the BMR signal is mostly dominated by the change in the Fermi surface convexity.

\end{abstract}

\maketitle 

The bilinear magnetoelectric resistance (BMR) effect is a powerful technique to extract important information on the band structure of quantum materials, such as Fermi surface convexity and spin textures \cite{He2018TI, He2018EG, Zhang2022, He2019, Liu2023}.
The BMR effect causes a modulation of the material's resistance depending on the relative angle between applied electric and magnetic fields \cite{SZhang2018}. 
This effect is also sometimes referred as the unidirectional magnetoresistance (UMR) \cite{Guillet2020, Calavalle2022} or electrical magnetochiral anisotropy \cite{Rikken2001, Wang2022, Yokouchi2023} and, as the name suggests, it has a linear dependence with both current and magnetic field.
The term UMR is also commonly used in literature to address the effect in magnetic materials \cite{Avci2015}. 
To avoid confusion, we will address this effect as BMR from now on.
This effect has been used to explore the spin-dependent Fermi surface of systems with different electronic properties \cite{He2018TI, He2018EG, Zhang2022}, and to give information on the shape and topology of the electron and hole pockets of semimetallic systems \cite{He2019}.

Weyl semimetals present an interesting platform to explore new physical phenomena due to their topologically protected states \cite{Burkov2016}.
These states, called Fermi arcs, appear at the surface of materials and connect the conduction and valence band  \cite{Deng2016}.
A lot of work has been done in understanding the influence of the topologically protected states on the transport properties of Weyl semimetals \cite{Chen2016, Qian2014, Zhang2018, Chen2018, YZhang2018, Khim2016, Sharma2017}, but a complete understanding of the role of bulk bands is still lacking.

\begin{figure*}[t!]
\includegraphics[width=\textwidth]{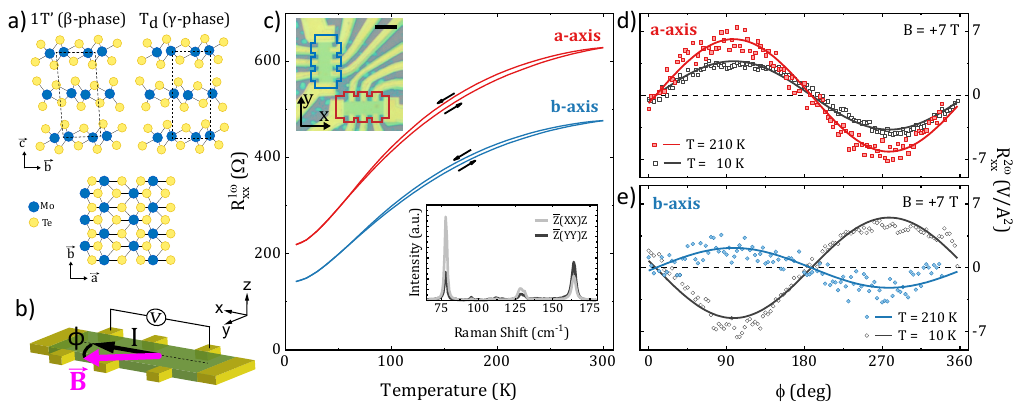}
\caption{\textbf{(a)} Crystal structure of semimetallic MoTe${}_2$ in the $1T'$ and $T_d$ crystal phases. \textbf{(b)} Schematics for the bilinear magnetoelectric resistance measurements. \textbf{(c)} Longitudinal resistance as a function of temperature measured with the current along the two main crystal axis of MoTe${}_2$. The top inset shows an optical image of the device (scale bar corresponds to \SI{5}{\micro\meter}) and the bottom inset shows the polarized Raman spectra for each axis. BMR measurements with the current applied along the a- \textbf{(d)} and b-axis \textbf{(e)} with an in-plane magnetic field of \SI{7}{\tesla} for two different temperatures (\SI{10}{\kelvin} and \SI{210}{\kelvin}). A vertical offset was removed for clarity.} 
\label{fig1}
\end{figure*}

The Weyl semimetal candidate MoTe${}_2$ is a van der Waals material showing low crystal symmetry \cite{Zhang2016, SChen2016, Stiehl2019} and strong spin-orbit coupling (SOC) \cite{Naylor2016, Cui2019}. 
Moreover, it undergoes a crystallographic phase transition as a function of temperature \cite{Clarke1978}.
MoTe${}_2$ crystallizes in the monoclinic $1T'$ phase at room temperature and transitions to the orthorrombic $T{}_d$ phase below \SI{240}{\kelvin} \cite{Clarke1978, Hughes1978} (Fig. \ref{fig1}a).
The crystallographic lattice shows a metallic zigzag chain along the $b$-axis, leading to a strong resistance anisotropy in the $ab$-plane for both phases \cite{Hughes1978}.
The phase transition also leads to a change on the crystal symmetry group, from space group $P2_{1}/m$ to space group $Pmn2_1$, allowing for additional spintronic phenomena to occur \cite{Stiehl2019, Roy2022, Tenzin2023}.
It is important to note that $1T'$-MoTe${}_2$ possesses inversion symmetry, which implies absence of band splitting and thus, a zero net (bulk) spin-texture. This results in a Fermi surface that is fully spin-degenerate. However, a non-zero spin-texture can appear at the surfaces, allowing for a "hidden" spin texture in the $1T'$ phase \cite{Liu2015, Yuan2013, Riley2014}.
Interestingly, the $T{}_d$ phase is predicted to be a type-II Weyl state with two Weyl nodes close to the Fermi level, while the $1T'$ phase remains a trivial semimetal \cite{Sun2015, Wang2016, Jiang2017}.
Although a large magnetoresistance has been measured in $T{}_d$-MoTe${}_2$ \cite{Chen2016, Lee2018, Pei2017}, the origins of the effect are not yet clear, since theoretical calculations indicate that MoTe${}_2$ is an uncompensated semimetal \cite{Thirupathaiah2017}.
Moreover, despite these interesting properties and promising spintronic \cite{Lim2018, Safeer2019, Stiehl2019, Zhou2019, Ontoso2023, Hoque2021} and topological applications \cite{Qian2014, Luo2016, YZhang2018, Zhang2018, Chen2018, Singh2020}, no study of its nonlinear magnetoresistance, i.e. the BMR effect, has been performed so far.

In this work, we report on the BMR effect observed for the two main crystal axis of semimetallic MoTe${}_2$ as a function of temperature. 
We observe a change in the BMR signal as we cool down from room temperature to \SI{10}{\kelvin} which is strongly anisotropic with respect to the crystal axes (Fig. \ref{fig1}).
For some of our devices we observe a crystal-phase change as a function of temperature, which apparently is partially correlated to the change in the BMR signal.
Despite this crystal phase change, density-functional theory (DFT) calculations show little changes on the overall band structure or band convexity at the Fermi level for both phases.
This indicates that the strong dependence of the BMR signal we observe could be attributed to a change on the Fermi surface convexity resulting from a Fermi level shift as a function of temperature, as observed in the sister material $T_d$-WTe${}_2$ \cite{He2019}.

Our devices consist of  $1T'$-MoTe${}_2$ encapsulated in hexagonal boron nitride (hBN), fabricated by mechanically exfoliated crystals (HQ Graphene), and stacked on top of each other via a dry van der Waals assembly technique \cite{Zomer2014}.
The hBN/$1T'$-MoTe${}_2$/hBN stack is assembled and placed on a SiO${}_2$/Si substrate under a nitrogen environment to protect the MoTe${}_2$ crystals from degrading.
Two perpendicular Hall bars were patterned on the stack, on the same MoTe${}_2$ flake, using electron-beam lithography and followed by Ti/Au contact deposition by conventional techniques.
The channel of each Hall bar was designed to be aligned with one of the main axis of the crystal.
The measurements were performed using conventional harmonic low-frequency (\SI{177}{\hertz}) lock-in techniques with current biases below $I_0 =$ \SI{500}{\micro\ampere} in a variable-temperature insert in Helium gas.
The in-plane angle ($\phi$) between an external magnetic field (up to $B =$ \SI{7}{\tesla}) and the current direction was varied while measuring the second-harmonic voltage in the longitudinal direction (Fig. \ref{fig1}b).
The measurements for the transverse direction can be found in the supplementary information \cite{supp}.
We have measured three different sets of devices with different MoTe${}_2$ thicknesses ($t$).
The change in the BMR signal as a function of temperature is clearly observed in two of our samples, $t =$ \SI{6}{\nano\meter} and \SI{12}{\nano\meter}.
Here we report our findings on the sample with $t = \SI{12}{\nano\meter}$ as obtained by atomic force microscopy.
More details on the device fabrication and the results for other samples can be found in the supplementary information in \cite{supp}.

The complete device is shown in the top inset of Fig. \ref{fig1}c, outlining each Hall bar with different colors to indicate the alignment of the channel with the $a$- (red) and $b$-axis (blue).
To confirm this alignment we performed polarized Raman spectroscopy measurements \cite{Zhou2017, Wang2017, Song2017} (bottom inset of Fig. \ref{fig1}c) and further characterized our sample by measuring the longitudinal (first-harmonic) resistance as function of temperature, as shown in Figure \ref{fig1}c.
As expected, a clear resistance anisotropy is observed between the two axis, where the low-resistive $b$-axis is along the metallic chain.
Moreover, a small hysteresis loop appears in the interval between \SIrange{80}{300}{\kelvin} as the sample is cooled down and warmed up again.
This behavior has been observed in thin MoTe${}_2$ flakes and confirmed to be due to a phase transition from the $1T'$ to the $T_d$ phase.
Different than for bulk MoTe${}_2$ this phase transition is not abrupt, resulting from a coexistence of the two phases, with the main crystal phase changing between the $1T'$ and the $T_d$ phases with temperature \cite{RHe2018, Cheon2021}.

The BMR effect arises from the interplay between the applied current and magnetic field.
Mathematically, we can express the resistance of a material by a resistance term R${}_0$, plus a term linear on the current and magnetic field: $\text{I} \text{R}_1\left( \mathbf{B} \right)$.
For an ac-current $\text{I} = \text{I}_0 \text{sin}(\omega t)$ applied through the material, therefore, the longitudinal voltage can be easily written by Ohm's law, yielding:
\begin{equation}
    \text{V} = \text{R}_0 \text{I}_0 \text{sin} \left( \omega t \right) + \frac{1}{2}\text{R}_1 \text{I}_0^2 + \frac{1}{2}\text{R}_1 \text{I}_0^2 \text{sin} \left( 2\omega t - \frac{\pi}{2} \right)
\end{equation}
\noindent with the first-harmonic, DC and second-harmonic components of the longitudinal voltage signal being the first, second and third terms on the right, respectively.
As can be seen, both DC and $\text{V}_{xx}^{2\omega}$ are proportional to the BMR coefficient $\text{R}_1$.
Here we focus on the second-harmonic response, which is less prone to additional artifacts.

We observe a remarkably different behavior on the second-harmonic longitudinal resistance ($\text{R}_{xx}^{2\omega}$) as a function of temperature for the two crystal axes.
Here we define $\text{R}_{xx}^{2\omega} = \frac{\text{V}_{xx}^{2\omega}}{(I_0)^2}$.
Figure \ref{fig1}d and e show the results for $I_0$ parallel to the $a$- and $b$-axis, respectively, for a magnetic field of +\SI{7}{\tesla} and two different temperatures (\SI{10}{\kelvin} and \SI{210}{\kelvin}).
Experimental data is represented by the scattered points, while the full lines are the corresponding fits according to:
\begin{equation}
    \text{R}_{xx}^{2\omega} = \Delta\text{R}_{xx}^{2\omega} \text{sin} [ 2 \left( \phi + \phi_0 \right) ] + y_0,
\end{equation}
\noindent where $\Delta\text{R}_{xx}^{2\omega}$ is the amplitude of the signal, and $\phi_0$ and $y_0$ are the angular and vertical offsets, respectively.
For both axes we notice that $\text{R}_{xx}^{2\omega}$ shows a sinusoidal behaviour with a periodicity of $2\pi$.
Regions of high and low resistance states can be found at angles of about \SI{90}{\degree} and \SI{180}{\degree}.
This agrees with a picture of a BMR signal arising from a Rashba-like spin-texture for the Fermi surface of MoTe${}_2$.
Strikingly, the two axes show very different behaviours as the temperature is reduced. 
For the $a$-axis, we observe only a small difference between the amplitudes at \SI{10}{\kelvin} and \SI{210}{\kelvin}, while for the $b$-axis the amplitude is not only $\sim2.5$ times larger at \SI{10}{\kelvin}, but it also changes sign compared to the measurement at \SI{210}{\kelvin}.
Further characterization of our signals, showing the magnetic field and current dependence of $\text{R}_{xx}^{2\omega}$ for both a- and b-axis can be found in the supplementary information \cite{supp}.

\begin{figure}[t!]
\includegraphics[width=\columnwidth]{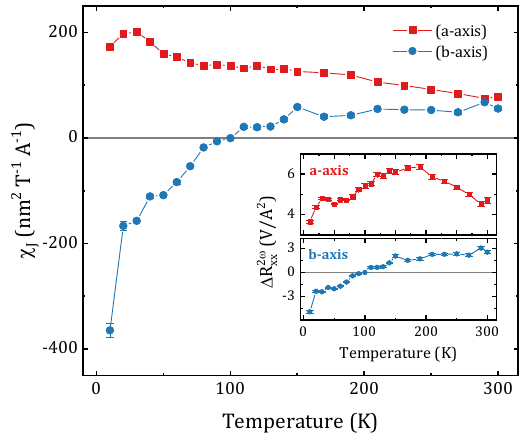}
\caption{The efficiency $\rchi_{J}$ for the current applied along the a- and b-axis as a function of temperature. The inset shows the amplitude $\Delta \text{R}_{xx}^{2\omega}$ of the fittings for the second-harmonic resistance measurements.} 
\label{fig2}
\end{figure}

\begin{figure*}[t!]
\includegraphics[width=\textwidth]{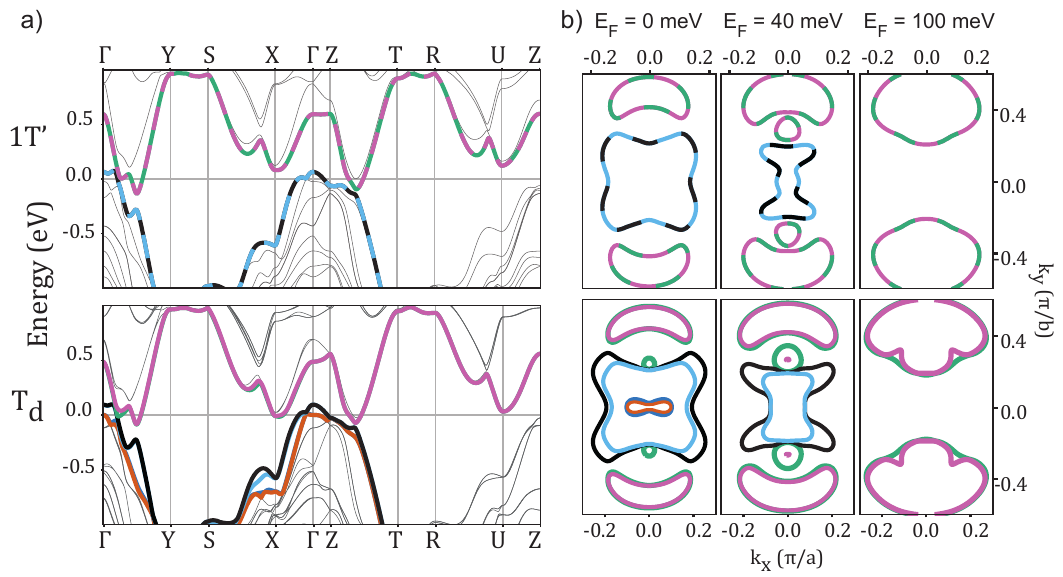}
\caption{\textbf{(a)} Electronic dispersion for the  $1T'$ and $T_d$ phases, and \textbf{(b)} corresponding Fermi surfaces at different energy cuts in respect to the conduction bands: $E_F = $ \SI{0}{\milli\electronvolt}, \SI{40}{\milli\electronvolt} and \SI{100}{\milli\electronvolt}. For the bands (and corresponding Fermi surfaces) of the $1T'$ phase, the mixing of two colours represents the spin degenerate bands.} 
\label{fig3}
\end{figure*}

In order to elucidate the origins of the different behavior for the two crystal axes, we perform similar measurements at various temperatures, inset of Fig. \ref{fig2}.
Interestingly, $\Delta \text{R}_{xx}^{2\omega}$ remains fairly constant for the $a$-axis, while for the $b$-axis the amplitude goes from positive to negative as the temperature decreases.
We note that $\Delta \text{R}_{xx}^{2\omega}$ can strongly depend on the specific device geometry and on the temperature dependence of the resistivity.
In order to exclude such effects, we calculate the BMR efficiency, defined as: 
\begin{equation}
    \rchi_{J} = \frac{2 \Delta \text{R}_{xx}^{2\omega} w t}{\text{R}_{0}\text{B}},
\end{equation}
\noindent with $w$ being the channel width.
The values for the BMR efficiency as a function of temperature are plotted in Figure \ref{fig2}.
We see that $\rchi_{J}$ reaches similar and positive values for both the $a$- and $b$- axes at high temperatures.
Since a spin-dependent BMR signal is not expected at the $1T'$-phase, a discussion on the potential causes for the observed BMR signals can be found in the supplementary information \cite{supp} (see also reference \cite{Beams2016} therein).
As the temperature decreases, the BMR efficiency increases modestly for the $a$-axis, peaking around \SI{30}{\kelvin}.
Differently, $\rchi_{J}$ decreases drastically for the $b$-axis, crossing zero at around \SI{100}{\kelvin}.
Cooling down further, the BMR efficiency drops rapidly for the $b$-axis and at \SI{10}{\kelvin} we obtain a BMR magnitude of twice as large as the one for the $a$-axis.
The computed values are $\rchi_{J}^{A}=$ \SI{173(3)}{\nano\meter\squared\per\tesla\per\ampere} and $\rchi_{J}^{B}=$ \SI{-364(13)}{\nano\meter\squared\per\tesla\per\ampere} for the $a-$ and $b-$axes, respectively.
These values and temperature behavior are comparable to the ones reported for the sister material WTe${}_2$ \cite{He2019}.
However, we point out that different from MoTe${}_2$, WTe${}_2$ does not undergo a crystal phase transition with a change in temperature.
The results for WTe${}_2$ were explained as a change in the convexity of the electron pockets as a function of the Fermi level \cite{He2019}.
Nonetheless, it is known that BMR can also depend on the spin texture of the Fermi surfaces \cite{He2018TI, He2018EG, Zhang2022}.

To further understand the origins of the behavior we observe, we performed DFT calculations to compute the Fermi contours of the hole and electron bands at different Fermi levels and for both crystal phases. Details of the calculations can be found in the supplementary material \cite{supp} (see also references \cite{giannozzi2009quantum, qe2, PAW1, PAW2, grimme2010consistent, PBE_GGA, paoflow1, paoflow2} therein).
As shown in Fig. {\ref{fig3}}a, the overall electronic dispersion presents some similar features for both $1T'$ and $T_d$ phases. 
Both show a semimetallic behavior, with electron and hole pockets at the Fermi level.
Due to the presence of inversion symmetry in the $1T'$ phase, we do not see any spin splitting in the DFT calculations. 
However, the $T_d$ lacks inversion symmetry and shows a Rashba-like spin texture, i.e. with the spins perpendicular to the crystal momentum (see supplementary material for the spin-dependent electronic dispersion).
For this reason, we rule out that the BMR signal is dominated by a spin-dependent contribution, since the second-harmonic signal reverses sign with temperature which would imply a reversal on the spin direction.

Similar to WTe${}_2$, it has been demonstrated that MoTe${}_2$ can show a shift on the Fermi level as a function of temperature \cite{Wu2015, Xu2018, Beaulieu2021, Kim2021}.
To explore the implications of this to our measurements, we obtain the Fermi-surfaces at different energies (Fig. {\ref{fig3}}) for both phases.
The overall behaviour of the electron and hole pockets with the change in energy is similar for both $1T'$ and $T_d$ MoTe${}_2$, however the latter shows two extra hole pockets appearing at low energies.
Remarkably, we obtain a change in convexity of the electron pockets for both phases along the $\Gamma - Y$ direction.
This change in convexity is consistent with the sign change we observe in our measurements, revealing that our measurements are dominated by a change in the Fermi surface convexity.

Our observations of a BMR signal dominated by the Fermi-surface convexity in $\text{MoTe}_{2}$ are an important step for the understanding of the band structure of this Weyl semimetal candidate.
Here we report an anisotropic behavior of the BMR signal as a function of temperature, with a clear inversion below \SI{100}{\kelvin} for the current along the $b$-axis while no sign reversal is observed for the current along the $a$-axis.
This observation is in agreement with a change in the Fermi surface convexity, induced by a change in the Fermi level due to temperature, similar to what has been reported for the sister material WTe$_{2}$ \cite{He2019}.
We envision that our demonstration of large BMR signals in $\text{MoTe}_{2}$ can be exploited in spintronic devices consisting of $\text{MoTe}_{2}$ interfaced with two-dimensional magnets, for which the magnetic field role could be played by the magnetic exchange. This would allow one to obtain important information on the magnetization direction even on magnetic insulators, while also exploiting the unusual spin torque symmetries \cite{Stiehl2019} provided by $\text{MoTe}_{2}$ for magnetization manipulation.

\section*{Data Availability}
The raw data and the data underlying the figures in the main text are publicly available through the data repository Zenodo at \url{https://doi.org/10.5281/zenodo.10592551}.

\section*{Acknowledgements}
We thank T. Liu for the valuable knowledge shared on measurements and sample fabrication.
We thank J.G. Holstein, H. Adema, H. de Vries, A. Joshua and F.H. van der Velde for their technical support.
This work was supported by the Dutch Research Council (NWO) through grants STU.019.014, the Zernike Institute for Advanced Materials, the research program “Materials for the Quantum Age” (QuMat, registrationnumber 024.005.006), which is part of the Gravitation program financed by the Dutch Ministry of Education, Culture and Science (OCW), and the European Union (ERC, 2DOPTOSPIN, 101076932).
Views and opinions expressed are however those of the author(s) only and do not necessarily reflect those of the European Union or the European Research Council.
Neither the European Union nor the granting authority can be held responsible for them.
The device fabrication and nanocharacterization were performed using Zernike NanoLabNL facilities.

\bibliographystyle{apsrev4-2}

\end{document}